# Influence of nitrogen on the growth of vertical graphene nanosheets under plasma

Subrata Ghosh[1,*,⊥], S. R. Polaki[1,*], Nanda Gopala Krishna[2], M. Kamruddin[1]

[1] *Surface and Nanoscience Division, Materials Science Group, Indira Gandhi Centre for Atomic Research, Kalpakkam - 603102, India.*
[2]*Corrosion Science and Technology Group, Indira Gandhi Centre for Atomic Research, Kalpakkam - 603102, India.*

**Abstract**

We have investigated the effect of nitrogen ($N_2$) as a carrier gas on the growth of vertical graphene nanosheets (VGN) by plasma enhanced chemical vapor deposition (PECVD). It is demonstrated that addition of nitrogen gas with a hydrocarbon precursor can enhance the nucleation and growth rate of graphitic base layer as well as vertical sheets. The gas also simultaneously acts as an etchant and a dopant element. The density of vertical sheets increases up to certain limit and start to decrease with further increase in $N_2$ concentration. The synthesized VGN exhibit sheet resistance from 0.89 to 1.89 KΩ/□ and mobility from 8.05 to 20.14 $cm^2$/V-s, depending on the morphology and type of carrier concentration. These results reveal that the surface morphology and electronic properties of VGN can be tuned by incorporation of nitrogen gas during the growth phase.

Corresponding author: Subrata Ghosh (subrataghosh.phys@gmail.com), S. R. Polaki (polaki@igcar.gov.in)

[⊥] Present affiliation of author: Department of Chemical Engineering, Chungbuk National University, 1 Chungdae-ro, Seowon-Gu, Cheongju, Chungbuk 28644, Republic of Korea

## 1. Introduction

Vertical graphene nanosheets (VGN), often known in literature as carbon nanowalls (CNW), graphitic petals or few-layered graphene nanoflakes, can be described as a well-controlled network of vertically standing, self-organized, interconnected few layer graphene (FLG) sheets [1]. These vertically standing sheets are the order of few nanometers thick, with length and height ranging from nanometer to micrometer. The incredible properties of VGN include high surface area, unique morphology and orientation, open edge structure, and remarkable optical, thermal, electrical and mechanical properties, easy functionalization, thermal and chemical stability [2-12]. These properties make VGN a material of research interest in a wide range of fields such as field emission, gas and bio sensor, energy applications, blackbody coating, spintronics and so forth [13-17]. In order to realize its potential usage in wide ranges of applications, it is necessary to control structure and properties [12]. It is thus a necessity to understand the growth mechanism towards achieving VGN structures with tailored properties.

Plasma enhanced chemical vapor deposition (PECVD) method is one of the widely used technique to synthesis VGN structures due to its inherent advantages such as a low synthesis temperature, lower activation energy (0.57 eV) requirement, catalyst-free and transfer free growth over a large area of substrate of choice etc. [18, 19]. Wide varieties of hydrocarbon precursors are extensively used for the growth of carbon materials using PECVD technique. Dilution of the hydrocarbon precursor with carrier gases such as Ar, $H_2$ is a crucial step in the growth of carbon materials [18-23]. In our recent study it has been observed that a flip in growth orientation of graphitic structure is achieved by simply changing the dilution gas composition (Ar/$H_2$) from Ar rich to $H_2$ rich [24]. While a Ar-rich gas composition resulted in growth of VGN structures, a $H_2$-rich gas composition resulted in growth of planar nanocrystalline graphite. In addition to gaseous Ar and $H_2$, nitrogen containing gas ($N_2$ or $NH_3$) too have been found to be a suitable carrier gas for the growth of carbon materials [22, 25-27]. The incorporation of $N_2$ into the carbon lattice is a possible route to modify the structural morphology, properties and electronic structure in carbon nanostructure [28, 29]. For instance, Shang *et al.* [22] reported VGN growth under $CH_4$/$N_2$ composition using PECVD and observed a higher growth rate of 96 μm/hr. Surprisingly, no nitrogen signature was observed from X-ray photoelectron spectroscopy (XPS) investigation. Cheng *et. al.* [23] through Optical emission spectroscopy studies established that introduction of $N_2$, increases the $C_2$ and CN radical density which in turn promote growth of VGN in PECVD. On the other hand, the $N_2$ also acts as a dopant for carbon nanostructures as reported by Takeuchi *et al.* wherein an electronic structural transformation of VGN from p-type (un-doped VGN) to n-type (doped VGN) was observed [29]. Moreover, $N_2$ is a known chemical etchant for amorphous carbon (*a*-C) and also performs better than the $H_2$ for carbon nanotube, diamond films and VGN [22, 30, 31]. However, no detailed report is available on the influence of $N_2$ in growth mechanism of VGN structures and its effect on the structural and electrical properties.

The principle focus of this study is to probe the influence of $N_2$ incorporation on the growth mechanism of VGN structures synthesized using Electron cyclotron resonance (ECR)–PECVD technique. The feedstock gas composition is varied to significantly influence the growth and structure of VGN samples. Substantial difference in growth mechanism of VGN structures due to the impurity existed in the carrier gas (3N pure Ar) compared to ultra-high pure (5N pure Ar) one is demonstrated. We also validated a role of nitrogen impurity in enhancing the VGN growth by mixing 5N pure $N_2$ with the carrier gas in a controlled manner. The correlation of structural quality and electronic properties brought out using Raman, XPS and four probe electrical measurements. Finally, we established a methodology to tailor morphology, structural quality and electronic properties of VGN structures to widen its applications.

## 2. Experimental methods
### 2.1 Growth of VGN by ECR-CVD

The VGN were grown on $SiO_2$/Si substrates using ECR–PECVD. To investigate the influence of $N_2$, 5N and 3N purity Ar, 5N purity $CH_4$ and $N_2$ were used wherein the 3N pure Ar gas known to contain a small amount of nitrogen as impurity. The feedstock gas composition used for the growth of VGN in this study is listed in Table 1. The details of the growth process are as reported in our previous study [32]. Briefly, the growth chamber is evacuated to $10^{-6}$ mbar pressure using turbomolecular pump backed by rotary pump, after loading the samples. Thereafter, the temperature of the heater was raised upto 800 $^0$C with ramping rate of 15 $^0$C/min. The substrates were annealed in Ar gas atmosphere for 30 minutes. The Ar plasma pretreatment was carried out with microwave power of 375 W to create dangling bonds on the substrates. These dangling bonds are effective to nucleate the growth of graphene films. Following the pre-treatment, the hydrocarbon source ($CH_4$) were fed into the chamber along with the desired career gas combination, as listed in Table 1. The growth was carried out for duration of 30 min at 800°C with microwave power of 375W. The gas molecules decompose into several neutral, positive species and electrons under plasma, which promotes the growth of graphitic films. As-grown films were subjected to annealing at the same temperature without plasma for better crystallinity [33]. Finally, the as-grown nanographitic films are cooled down naturally and taken out to investigate the influence of $N_2$ gas on the growth.

Table1 – Gas composition for the growth of VGN in PECVD

| Sample | Gas Composition | | |
|---|---|---|---|
| | $CH_4$ [sccm] | Ar[sccm] | $N_2$[sccm] |
| S1 | 5[5N] | 25 [5N] | 0 |
| S2 | 5[5N] | 25 [3N] | - |
| S3 | 5[5N] | 20 [5N] | 5[5N] |
| S4 | 5[5N] | 20 [3N] | 5[5N] |
| S5 | 5[5N] | 0 | 25[5N] |

**2.2 Characterization**
Field Emission Scanning Electron Microscope (FESEM, Supra 55, Zeiss) was employed to obtain the morphology of the VGN structures. The microstructure of as-grown film was evaluated by high-resolution transmission electron microscope (HRTEM; LIBRA 200FE, Zeiss). The structural properties, in terms of crystallinity, defects and disorder, were evaluated by micro-Raman spectroscopy (Renishaw inVia, UK) with 514 nm excitation and 100× objective lens. In order to avoid the laser induced heating of sample, the laser power was maintained below 1 mW. Chemical structure and bonding were evaluated using XPS (M/s SPECS Surface Nano Analysis GmbH, Germany) with a monochromatic Al Kα radiation (Λ = 1486.71 eV). Data was processed by Specslab2 software and analyzed using Casa XPS software, which can deconvolute each spectrum using a pseudo-voigt function after the shirley background subtraction with best fitting parameters ($R^2$ = 0.9998). Sheet resistance and Hall measurements were performed on the all samples using Agilent B2902A precision source/measure unit in van-der Paw geometry. The electrical contacts were prepared using a conductive silver paste.

## 3. Results and discussions
### *3.1 Morphological analysis*
The surface morphology of the VGN grown on $SiO_2$/Si substrates under different gas composition are illustrated in Fig. 1(a-e). A significant variation in growth is evident from the FESEM imaging. A well separated free-standing VGN and nanograhitic (NG) base layer for the S1 is observed, where 5N purity Ar is used as carrier gas (Fig. 1a). Figure 1(b) shows the morphology of the sample S2 grown with 3N pure Ar carrier gas, a well grown VGN structures with higher number density are observed in S2. The wall height is measured to be 150 nm, as showed in the corresponding cross section SEM image [Fig. 1(f)]. Noteworthy to mention, the thickness of NG layer is found to be lower than that of S1. The significant change in morphology can be attributed to the variation in purity of carrier gases, as it is the only variable. To substantiate this, we chose to use the mixture of Ar and $N_2$ gases (20 and 5 sccm, respectively) as carrier gas instead of Ar alone, details are given in the Table 1. The effect of additional nitrogen to the carrier gas on the morphology is depicted in Fig. 1(c) and (d) corresponding to the sample S3 and S4, respectively. In both cases, the growth is contrasting to their corresponding counter parts Fig. 1(a) and 1(b), respectively. The significant role of nitrogen presence on the VGN growth is clearly observed. Even though the vertical height of the VGN is almost same (146 nm) for S3 (Fig. 1g), the structure is observed to have lower density and lower vertical alignment. Increase in disorder and curvature are observed in the vertical sheets of S3 (Fig. 1c) which can be attributed to the incorporation of $N_2$ [34]. In the case of N2 mixed with 3N pure Ar gas used as a carrier gas, isolated VGN growth is observed. To further demonstrate the effect of nitrogen inclusion on the growth of VGN structures, we used 5N pure $N_2$ as the sole carrier gas. The corresponding morphology, showed in Fig. 1(e) illustrates the occurrence of isolated NG island growth instead of VGN. The observed results indicate that inclusion of slight amount $N_2$ promotes the growth of VGN, however, excess amount of $N_2$ etches the vertical sheets as well as the base layer too. This is well corroborated with the higher growth rate exhibited by sample S2, grown under 3N pure Ar carrier gas.

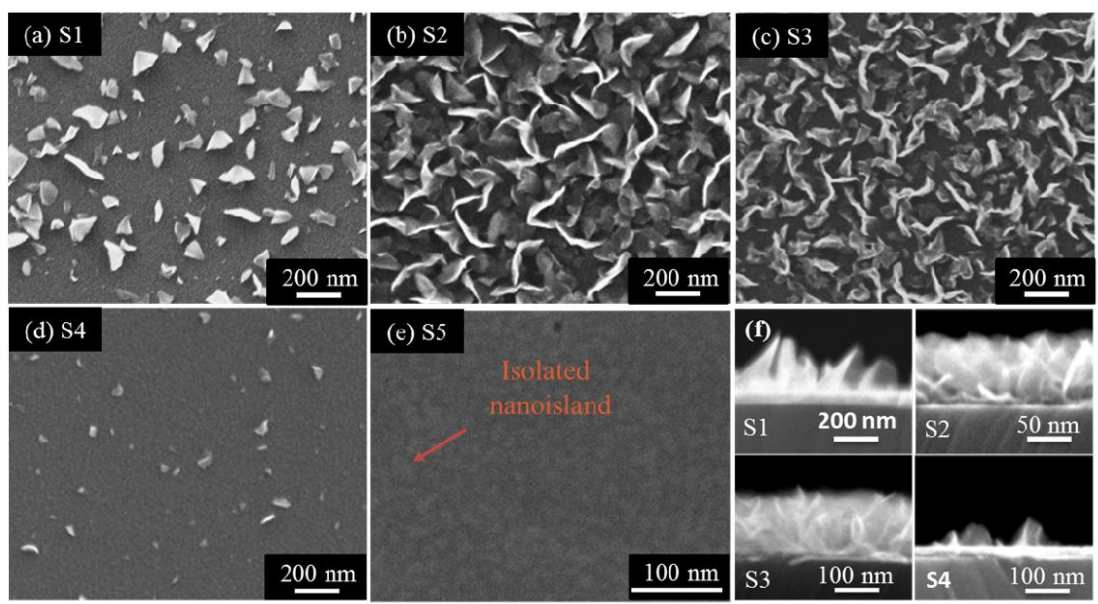

Fig .1: Top view of SEM micrographs of (a) S1, (b) S2, (c) S3, (d) S4, and (e) S5. (f) the cross-section SEM image of the for respective VGN

Figure 2 illustrates the TEM micrographs of the nanographitic films. Samples S2 and S4 has been chosen to analyse the microstructure. Each vertical sheet contains 3-10 layers of graphene with interlayer spacing of 0.365 nm, as sheen from Fig. 2(a). Fig. 2(b) reveals that the each nanographitic islands consists of 46 graphene layers with spacing 0.386 nm. The larger interlayer spacing than that of graphite (0.334 nm) confirms its turbostatic nature [24].

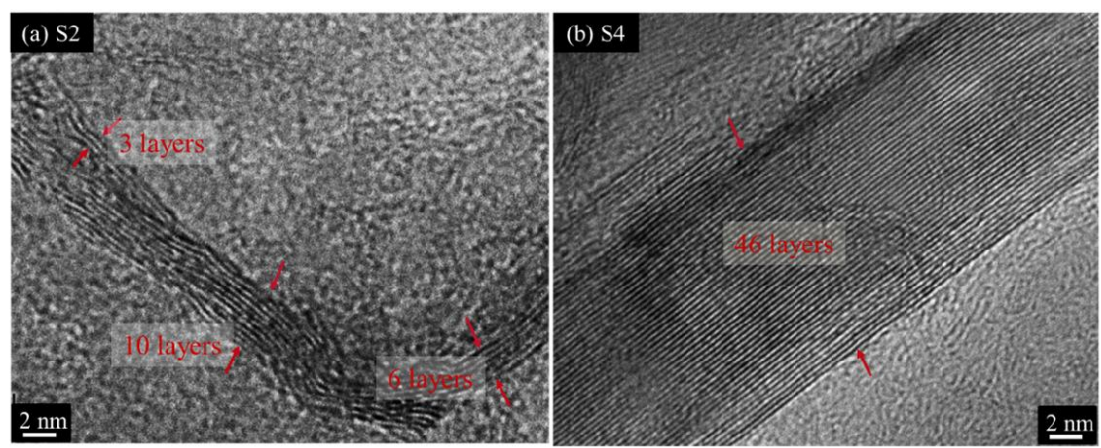

Figure 2: High resolution transmission electron micrographs of (a) S2 and (b) S4

The above observation can be explained based on notion that the introduction of small amount of N$_2$ increases the C$_2$ and CN radical density [equation 1 and 2] as described by Cheng *et. al.* [23], in addition to atomic N$_2$ partially etching the *a*-C. However further introduction of N$_2$ is observed to not favor VGN growth, resulting in isolated NG islands as exhibited by the sample S5. This can be attributed to the increase in density of CN radicals in comparison to that of C$_2$ radicals which in turn suppresses the C$_2$ radical movement into the substrate [23]. From equation 2, it is also clear that higher N$_2$ inclusion during the growth leads to higher amount of atomic hydrogen along with atomic nitrogen species [23], leading to reduction in C$_2$ radical density due to recombination process. This also acts as an effective etchant of the carbon species, thus resulting in a lower growth rate. Although, *in-built* local electric field presented in plasma, after graphitic base-layer formation, vertical growth is restricted. In case of S5 where only N$_2$ is used as carrier gas, a lot of isolated NG island formation taken place.

$$CH_4 \xrightarrow{Ar} CH_x\ (x = 0-3) \xrightarrow{Ar} C_2H_2 \xrightarrow{Ar} C_2 \qquad (1)$$

$$CH_P + N_2^* \rightarrow CN + N + H_q + H_r (p = q + r) \qquad (2)$$

### 3.2 Raman spectroscopic analysis

Raman spectroscopy is an excellent highly sensitive and non-destructive technique to investigate the structural and electronic properties of pristine as well as doped carbon nanostructures. The Raman spectra consists of prominent bands such as D, G, D′, G′, D+D′ which are typical for a defected graphitic system [21]. While the G- and G′- band confirm the presence of graphitic structure, D-, D′- and D+D′- bands are defect related peaks of carbon nanostructure that arise due to disorder in structure and presence of doping element [21]. Figure 3 shows the Raman spectra of the VGN grown under different gas compositions. The Raman spectra are further deconvoluted and all the extracted parameters are listed in table 2. Usually, the FWHM in the carbon structures are dependent on number of graphene layers edge density, defects, strain, doping and so on [35]. In this study, the change in FWHM of graphene films are contributed from both NG layer and VGN. Although sheet density increases in S2 compared to S1 (Fig. 1 a and b), the reduction in FWHM can be attributed to the improved crystallinity. The width (FWHM) of both the D and G bands are observed to increasing nitrogen to the Ar gas irrespective of its purity (5N or 3N). The initial increased FWHM in S3 compared to S2 may attributed to the curvatures in each vertical sheet, as seen from Fig. 1(c). Further increase in FWHM in the S4 and S5 are due to the presence of more number of graphene layers (Fig. 2b), many NG island, which contains C-H species, more NG grain boundaries (Fig. 1 d and e). The result also signifies the reduction in crystallite size with nitrogen introduction. In addition, it is The increased FWHM is also attributed to alteration in chemical bonding locally due to nitrogen introduction. The G peak is seen to shift to higher wave number beyond 1580 cm$^{-1}$ with nitrogen incorporation, which signifies an increase in disorder, which is well corroborated with the increase in FWHM as discussed above. To validate these finding XPS analysis was carried and described in the following paragraph.

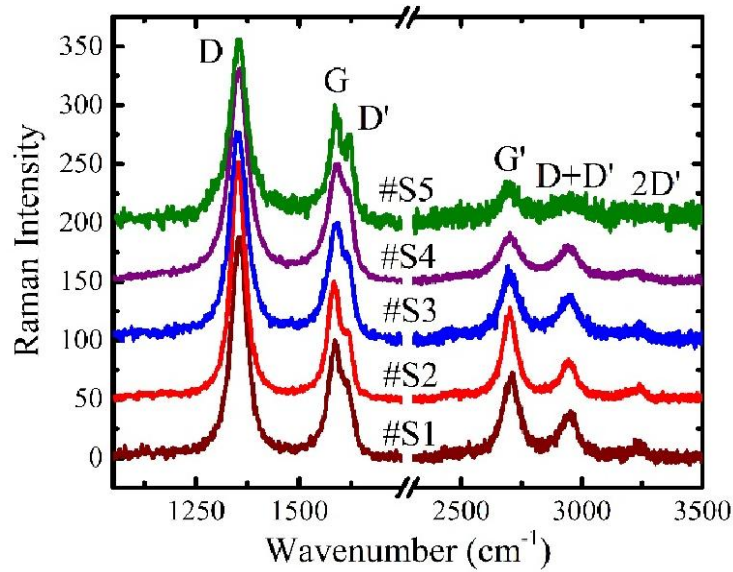

Figure 3: Raman spectra of the nanographitic film grown under different gas composition.

Table. 2: Extracted Raman parameters of VGNs grown under different gas composition

| sample | Peak position [cm$^{-1}$] | | | FWHM [cm$^{-1}$] | | |
|---|---|---|---|---|---|---|
| | D-band | G-band | G′-band | D-band | G-band | G′-band |
| #S1 | 1354.67 | 1586.92 | 2705.20 | 46.3 | 42.7 | 84.1 |
| #S2 | 1351.85 | 1583.82 | 2699.29 | 41.7 | 37.7 | 74.6 |
| #S3 | 1353.00 | 1589.32 | 2699.68 | 49.8 | 51.4 | 100.9 |
| #S4 | 1353.79 | 1591.25 | 2701.50 | 54.5 | 56.5 | 106.9 |
| #S5 | 1353.32 | 1589.52 | 2699.06 | 49.8 | 43.6 | 89.2 |

### 3.3 XPS analysis

XPS is an excellent surface analytical techniques for the family of carbon based materials in order to probe the elemental compositional and chemical bonding [36, 37]. Herein, we analyze the XPS spectra of the VGN grown under the compositions as listed in Table 1. Figure 4(a) depicts the XPS spectra (0 – 1200 eV) of all the VGN samples. No peak that corresponds to nitrogen is observed in S3 and S4 even though additional nitrogen was incorporated during the growth. This could be due to the fact that the concentration of nitrogen used was well below the detection limits in XPS technique. The sample S5 grown only with $N_2$ as the carrier gas exhibits the N1$s$ peak around 400 eV. From the SEM image (Fig. 1e), we have seen that the S5 sample contains isolated NG island. The Si2$p$ and high intense O1$s$ peak are from the exposed regions of underlying $SiO_2$ substrate. The C1s peak is deconvoluted to extract information on the chemical bonding state.

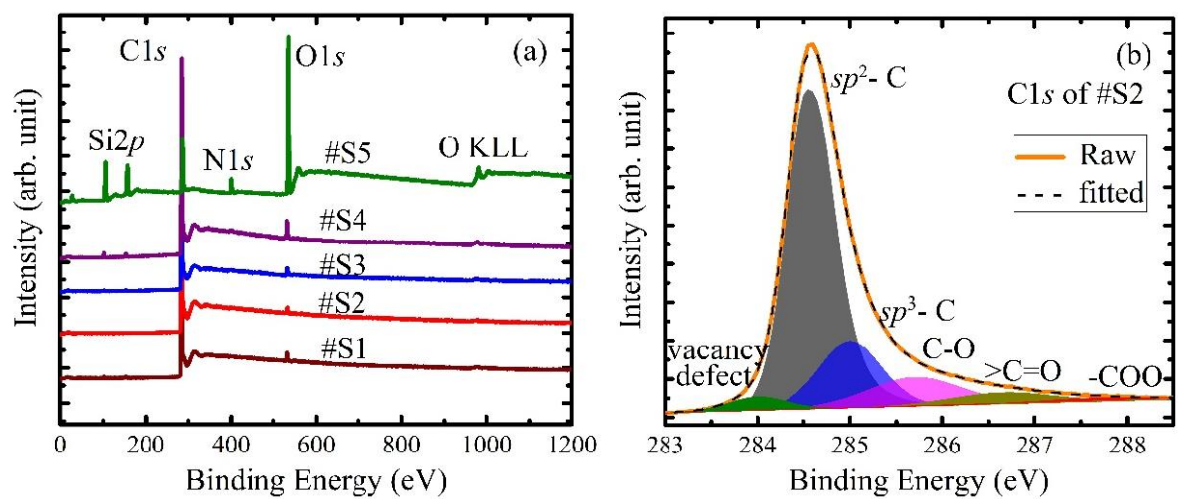

Fig. 4 (a) XPS survey spectra of nano-graphitic films and (b) C1*s* spectra of S2 sample

Figure 4(b) shows the high resolution deconvoluted C1*s* spectrum of the S2 sample. The C1*s* spectra shows a clear asymmetry in the higher binding energy (B.E.) regime for all samples. The line shape of C1*s* peak could be influenced by various types of defects such as $sp^2$ C-H and $sp^3$ C-H group, vacancies, Stone-Wales defects and functional groups attached with carbon atoms [37]. The C1*s* spectrum exhibits a predominant peak at around 284.54 eV, which corresponds to $sp^2$ C=C bonds and low intensity peaks at around 285, 285.7, 286.65 and 287.7 eV are attributed to the C-H $sp^3$, C-O, C=O and HO-C=O, respectively. The origin of these impurity peaks can be attributed to adsorbed moisture while exposing the samples to ambient atmosphere. The fitted parameters of C1*s* XPS spectra for different samples are listed in table 2. As shown in table 2, the $sp^2$C=C content of the film is highest upon dilution with Ar alone and low with introduction of additional $N_2$ during growth. Moreover, the $sp^2$ content is found to be lower in case of S2 than S1. It shows that the presence of even a very low value of nitrogen imurityt in 3N pure Ar carrier gas used to grow S2 causes reduction in $sp^2$ C=C. Interestingly, nitrogen incorporation increases the $sp^3$ C-C, thus the increase in disorder, as observed by Raman spectroscopy results. Apart from this, a peak near B.E of 284.0 eV is found in all grown samples which is assigned as defect peak. The defect peak could arise from vacancies, pentagon–heptagon rings, presence of $sp^2$ C-H group and other functional groups attached to edges of basal plane of graphitic lattice [37]. Overall, the percentage of defects is found to be increased upon incorporation of nitrogen gas during growth. The area under the defect peak is lower for S2 compared to other samples. Further, this defect peak could be corroborated with the fullerene or carbon onion-like structure formation at the early stage nucleation process of carbon nanostructures [38]. The XPS results are found to be well corroborated with the Raman spectroscopy analysis.

Table 3: Extracted parameters of XPS fitting of VGN grown under different gas composition

|     | C=C sp² | | C-H sp³ | | C-OH | | C=O | | HO-C=O | | defect | |
| --- | --- | --- | --- | --- | --- | --- | --- | --- | --- | --- | --- | --- |
|     | B.E. [eV] | Area [%] | B.E. [eV] | Area [%] | B.E. [eV] | Area [%] | B.E. [eV] | Area [%] | B.E. [eV] | Area [%] | B.E. [eV] | Area [%] |
| #S1 | 284.45 | 68.49 | 284.90 | 13.59 | 285.60 | 8.15  | 286.45 | 4.07 | 287.35 | 1.63 | 283.95 | 4.08 |
| #S2 | 284.54 | 63.33 | 285.00 | 16.59 | 285.70 | 10.80 | 286.65 | 4.58 | 287.70 | 1.53 | 284.00 | 3.17 |
| #S3 | 284.61 | 59.27 | 285.10 | 17.66 | 285.80 | 11.96 | 286.70 | 3.99 | 287.60 | 1.57 | 284.10 | 5.56 |
| #S4 | 284.58 | 53.16 | 285.00 | 19.47 | 285.70 | 10.99 | 286.65 | 5.61 | 287.90 | 2.17 | 284.10 | 8.59 |
| #S5 | 284.60 | 60.28 | 285.15 | 16.37 | 286.00 | 7.59  | 286.95 | 3.72 | 288.10 | 2.08 | 284.00 | 7.74 |

### *3.4 Electrical characterization*

Once the influence of N$_2$ on morphology and structural quality of VGNs have been investigated, it is essential to know its role on the electronic properties of the samples. To understand the influence of nitrogen incorporation on the electrical and electronic properties of VGN structures, we measure the sheet resistance, electrical and carrier concentration, mobility using Hall measurements in four-probe configuration. The Ohmic nature of the studied films is assured from current-voltage linear relationship (Fig. 5a). The electronic mobility and career concentration were calculated using the values of sheet resistance and slope of resistance versus magnetic field plot in Hall configuration (Fig. 5b and c). The parameters of electric measurements by four probe method are listed in Table 3. The measurements were carried out in the van der Pauw geometry on the basic assumption that all the samples are thin film in macroscale. S1 having lower number of vertical sheets than the S2, as evident from FESEM analysis, exhibits lower sheet resistance and higher mobility This can be explained from the FESEM and Raman spectroscopic analysis that the mobility and electrical resistance are depends on the quality of graphitic layers and number density of vertical sheets. Both the samples show *p*-type conductivity. In the case of S2, even ppm levels of N$_2$ in the carrier gas viz. 3N pure Ar, doesn't nudge it towards n-type conduction, which can be attributed to the role played by the size and rearrangement of neighboring C atoms [39]. Samples S3 and S4 exhibit *n*-type behavior as observed from the room temperature Hall measurement. Although S4 possess lower sheet resistance than S3 yet its mobility is low which can be attributed to poor degree of graphitization and lesser number density of sheets than S3. Sample S5 does not show any electrical continuity, that can be attributed to the discontinued NG island structure, as evident from the FESEM image shown in Fig. 1 (e).

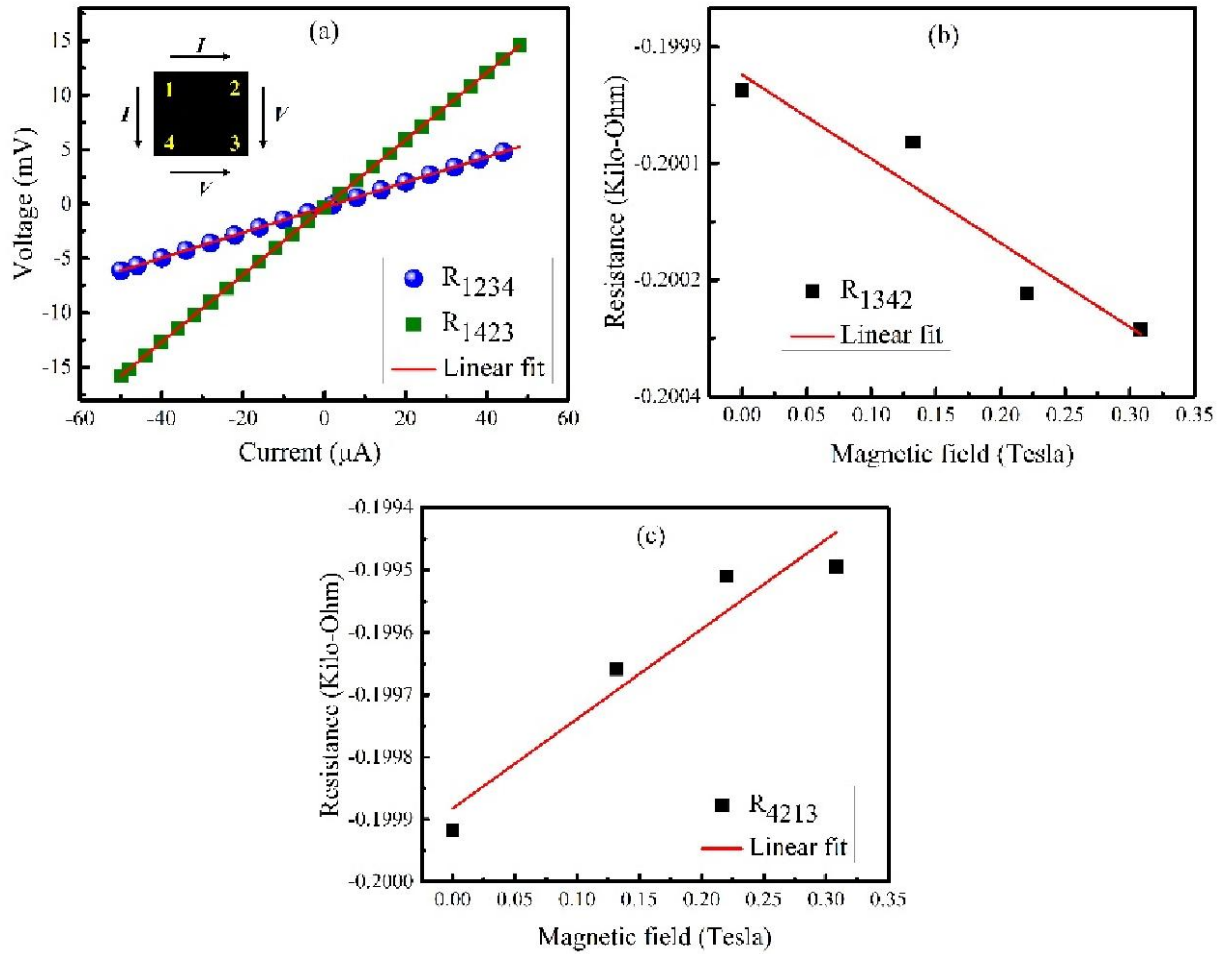

Figure 5: (a) I-V characteristics of nanographitic film, (b-c) resistance vs magnetic field plot for the nanographitic film in two configurations.

Table 4: Electrical parameters of VGN grown under different gas composition

| sample | Vertical height nm | Sheet Rest. KΩ/□ | Mobility $cm^{-2}/V\text{-}S$ | Carr. conc. x $10^{14} cm^{-2}$ | Type of carr. |
|---|---|---|---|---|---|
| #S1 | 120±11 | 0.89 | 14.14 | 4.96 | p |
| #S2 | 150±5 | 1.88 | 8.05 | 4.12 | p |
| #S3 | 147±3 | 1.89 | 20.14 | 1.64 | n |
| #S4 | 50±5 | 1.51 | 10.47 | 3.94 | n |
| #S5 | - | - | - | - | - |

## 5. Conclusion

We investigated the effect of $N_2$ inclusion on the growth mechanism of VGN structures. Inclusion of slight $N_2$ along with carrier gas was found to enhance the VGN growth rate. However, an excess amount of $N_2$ resulted in chemical etching and ended up with discontinuous NG island growth. Additional incorporation of nitrogen into the VGN structure was found to increase the defect density. Furthermore, sheet resistance is observed to increase with nitrogen incorporation. Overall, $N_2$ incorporation is found to significantly influence the structural quality and electronic properties of VGN structures. This study paves way for controlled architecting of VGN structure with suitable morphology, structural quality and desired electrical properties for its potential utilization in variety of applications.

## Acknowledgement

The authors thank S. Dhara for allowing access to the Raman spectroscopic facility, K. Ganesan for room temperature Hall measurement facility and S. Amirthapandian for TEM measurement. We are grateful to T. R. Devidas for careful evaluation of the manuscript. The authors also thank thank G. Amarendra for his kind support.

## References

[1] M Hiramatsu, M Hori (2010) Carbon nanowalls: synthesis and emerging applications. Springer,

[2] S Ghosh, G Sahoo, SR Polaki, NG Krishna, M Kamruddin, T Mathews (2017) J. Appl. Phys. 122: 214902. Doi:10.1063/1.5002748

[3] AK Sivadasan, P Santanu, G Subrata, P Ramanathaswamy, D Sandip (2017) Nanotechnology 28: 465703.

[4] S Vizireanu, G Dinescu, LC Nistor, et al. (2013) Physica E: Low dimens. Syst. Nanostruct. 47: 59. Doi:10.1016/j.physe.2012.09.027

[5] S Ghosh, SR Polaki, P Ajikumar, NG Krishna, M Kamruddin (2017) Ind. J. Phys. Doi:10.1007/s12648-017-1113-0

[6] K Davami, Y Jiang, J Cortes, et al. (2016) Nanotechnology 27: 155701. Doi:10.1088/0957-4484/27/15/155701

[7] S Ghosh, B Gupta, K Ganesan, et al. (2016) Mater Today Proc. 3: 1686. Doi:http://dx.doi.org/10.1016/j.matpr.2016.04.060

[8] KK Mishra, S Ghosh, RR Thoguluva, S Amirthapandian, M Kamruddin (2016) J. Phys. Chem. C 120: 25092. Doi:10.1021/acs.jpcc.6b08754

[9] S Ghosh, K Ganesan, SR Polaki, AK Sivadasan, M Kamruddin, AK Tyagi (2016) Adv. Sci. Eng. Med. 8: 146. Doi:https://doi.org/10.1166/asem.2016.1826

[10] Q Liao, N Li, S Jin, G Yang, C Wang (2015) ACS nano 9: 5310.

[11] S Evlashin, S Svyakhovskiy, N Suetin, et al. (2014) Carbon 70: 111. Doi:10.1016/j.carbon.2013.12.079

[12] HJ Cho, H Kondo, K Ishikawa, M Sekine, M Hiramatsu, M Hori (2014) Carbon 68: 380. Doi:10.1016/j.carbon.2013.11.014

[13] S Ghosh, T Mathews, B Gupta, A Das, NG Krishna, M Kamruddin (2017) Nano-Struct. Nano-Objects 10: 42. Doi:http://dx.doi.org/10.1016/j.nanoso.2017.03.008

[14] D Seo, A Rider, S Kumar, L Randeniya, K Ostrikov (2013) Carbon 60: 221. Doi:10.1016/j.carbon.2013.04.015

[15] V Krivchenko, S Evlashin, K Mironovich, et al. (2013) Sci. Rep. 3: 3328. Doi:10.1038/srep03328

[16] A Malesevic, R Kemps, A Vanhulsel, MP Chowdhury, A Volodin, C Van Haesendonck (2008) J. Appl. Phys. 104: 084301. Doi:10.1063/1.2999636

[17] SC Ray, N Soin, T Makgato, et al. (2014) Sci. Rep. 4: 3862. Doi:10.1038/srep03862

[18] Z Bo, Y Yang, J Chen, K Yu, J Yan, K Cen (2013) Nanoscale 5: 5180. Doi:10.1039/C3NR33449J

[19] S Ghosh, SR Polaki, N Kumar, S Amirthapandian, M Kamruddin, K Ostrikov (2017) Beilstein J. Nanotechnol. 8: 1658. Doi:10.3762/bjnano.8.166

[20] Y Wu, P Qiao, T Chong, Z Shen (2002) Adv. Mater. 14: 64. Doi:10.1002/1521-4095(20020104)14:1<64::AID-ADMA64>3.0.CO;2-G

Journal ref: *Journal of Materials Science*, **53**, 7316–7325 (2018) https://doi.org/10.1007/s10853-018-2080-3

[21]    S Ghosh, K Ganesan, SR Polaki, et al. (2014) J. Raman. Spectrosc. 45: 642. Doi:10.1002/jrs.4530
[22]    NG Shang, P Papakonstantinou, M McMullan, et al. (2008) Adv. Funct. Mater. 18: 3506. Doi:10.1002/adfm.200800951
[23]    C Cheng, K Teii (2012) IEEE Trans. Plasma Sci. 40: 1783. Doi:10.1109/TPS.2012.2198487
[24]    S Ghosh, K Ganesan, SR Polaki, et al. (2015) RSC Adv. 5: 91922. Doi:10.1039/C5RA20820C
[25]    AT Chuang, J Robertson, BO Boskovic, KK Koziol (2007) Appl. Phys. Lett. 90: 123107. Doi:10.1063/1.2715441
[26]    E Luais, M Boujtita, A Gohier, et al. (2009) Appl. Phys. Lett. 95: 014104. Doi:10.1063/1.3170033
[27]    S Vizireanu, L Nistor, M Haupt, V Katzenmaier, C Oehr, G Dinescu (2008) Plasma Process. Polym. 5: 263. Doi:10.1002/ppap.200700120
[28]    N Soin, SC Ray, S Sarma, et al. (2017) J. Phys. Chem. C 121: 14073. Doi:10.1021/acs.jpcc.7b01645
[29]    W Takeuchi, M Ura, M Hiramatsu, Y Tokuda, H Kano, M Hori (2008) Appl. Phys. Lett. 92: 213103. Doi:10.1063/1.2936850
[30]    T Vandevelde, T-D Wu, C Quaeyhaegens, J Vlekken, M D'Olieslaeger, L Stals (1999) Thin Solid Films 340: 159. Doi:10.1016/S0040-6090(98)01410-2
[31]    E Wang, Z Guo, J Ma, et al. (2003) Carbon 41: 1827. Doi:10.1016/S0008-6223(03)00155-6
[32]    S Ghosh, K Ganesan, S Polaki, et al. (2015) Appl. Surf. Sci. 349: 576. Doi:https://doi.org/10.1016/j.apsusc.2015.05.038
[33]    G Sahoo, S Ghosh, SR Polaki, T Mathews, M Kamruddin (2017) Nanotechnology 28: 415702. Doi:https://doi.org/10.1088/1361-6528/aa8252
[34]    J Mandumpal, S Gemming, G Seifert (2007) Chem. Phys. Lett. 447: 115. Doi:10.1016/j.cplett.2007.09.007
[35]    SS Nanda, MJ Kim, KS Yeom, SSA An, H Ju, DK Yi (2016) TrAC Trends in Analytical Chemistry 80: 125. Doi:https://doi.org/10.1016/j.trac.2016.02.024
[36]    G Karuppiah, S Ghosh, NG Krishna, S Ilango, M Kamruddin, AK Tyagi (2016) Phys. Chem. Chem. Phys. 18: 22160. Doi:10.1039/C6CP02033J
[37]    Y Yamada, H Yasuda, K Murota, M Nakamura, T Sodesawa, S Sato (2013) Journal of Materials Science 48: 8171. Doi:10.1007/s10853-013-7630-0
[38]    J Zhao, M Shaygan, J Eckert, M Meyyappan, MH Rümmeli (2014) Nano Lett. 14: 3064. Doi:10.1021/nl501039c
[39]    P Ayala, R Arenal, M Rümmeli, A Rubio, T Pichler (2010) Carbon 48: 575. Doi:10.1016/j.carbon.2009.10.009